\documentclass[aps,reprint,twocolumn,longbibliography,superscriptaddress]{revtex4-2}
\usepackage[english]{babel}
\usepackage{amsmath,amsfonts,amsthm,bm,braket}
\usepackage{graphicx}
\usepackage[colorlinks=true, allcolors=blue,breaklinks]{hyperref}
\usepackage{float}
\usepackage{physics}
\usepackage{tikz}
\usepackage[normalem]{ulem}
\usetikzlibrary{trees}
\usepackage{bbold}
\usepackage{comment}
\usepackage{adjustbox}
\usepackage{mathtools}
\usepackage{enumerate}

\begin{document}

\title{Measurement-induced asymmetry in bipartite networks}
\author{A. Lowe}
\email{adam.j.lowe90@gmail.com}
\affiliation{Aston University, Department of Applied Mathematics and Data Science, Birmingham, B4 7ET, UK}
\author{E. Medina-Guerra}
\affiliation{Department of Condensed Matter Physics, Weizmann Institute of Science, Rehovot 7610001, Israel}

\date{\today}

\begin{abstract}
We consider an interacting bipartite network through a Bayesian game-theoretic framework and demonstrate that weak measurements introduce an inherent asymmetry that is not present when using standard projective measurements. These asymmetries are revealed in the expected payoff for a Bayesian version of the Prisoner's dilemma, demonstrating that certain advantages can be found for given subsystems depending on the measurements performed in the network. We demonstrate that this setup allows measurement-induced control for one of the respective parties.
\end{abstract}

\maketitle

\section{Introduction}
Measurement theory is an increasingly active area of research, from metrology \cite{G_2014,PhysRevLett.112.210401,10.1063/10.0010204,StatEnt}, to understanding the fundamental nature of systems \cite{RevModPhys.38.453,physInt}. However, when considering quantum systems, the standard approach of using projective measurements only reveals certain aspects of the model due to the huge disturbance (or back action) they produce. To overcome this problem, weak measurements \cite{PhysRevLett.95.110409} are used as their back action can be small enough to perturb the measured system. As a trade-off, little information about the system is gained. 
More precisely, performing a weak measurement aims to extract a partial amount of information from a state without entirely reducing it to an eigenstate of a given observable. Weak measurements have been implemented to prepare quantum states \cite{Yuval1,morales_engineering_2023,Parveen1}, improve the efficiency in state tomography \cite{wu_state_2013,PhysRevLett.117.120401}, simultaneously measure non-commuting observables \cite{hofmann_sequential_2014,ochoa_simultaneous_2018,Chen:19}, reveal more correlation in multipartite systems \cite{SINGH2014141},  and to amplify signals in quantum metrology \cite{PhysRevLett.128.040503,torres_weak_2016,song_enhanced_2021} utilizing the closely related concept of \emph{weak-values} \cite{vaidman_weak_2009}. These generalize the concept of expectation values in quantum mechanics, among other applications \cite{PRXQuantum.4.020347}.

Alongside this, the understanding of interacting parties in a network \cite{doi:10.1126/science.286.5439.509, strogatz,doi:10.1073/pnas.122653799,ZHOU202057} is highly relevant across multiple scientific fields due to the increasing number of emerging interacting neworks. A host of different techniques are used to understand the underlying networks depending on its topology \cite{10.5555/3455716.3455900,NetTop}.  Understanding how each party can affect the overall system with their local measurements and operations is vital to understanding the overall nature of the system and the correlation between the parties \cite{Schmid2020typeindependent,8578911}.

A natural framework to understand and model how parties can interact is through game-theoretic techniques \cite{75b5cec4-4f40-3f36-9492-860b8376add8,540b73bd-a3f1-333e-a206-c24d0fbbb8bc}. This results in a scenario where the two parties perform certain operations and measurements in order to optimize their respective situation. These types of scenarios occur in everyday life, from buying and selling goods to determining the best route of a journey \cite{PhysRevX.11.011048,PhysRevLett.125.168301,https://doi.org/10.1049/qtc2.12066}, among many others.

It is well established that asymmetry can be present in correlations between interacting parties, either through, e.g., quantum discord \cite{PhysRevLett.88.017901,L_Henderson_2001}, or standard classical correlation. Asymmetry in this context 
relates to the asymmetric correlation between two parties. For example, the quantum discord of a given state can differ depending on which subsystem is measured and optimized. Despite this non-trivial aspect, understanding the link between measurement theory and asymmetry has received little attention \cite{PhysRevLett.127.170401,Franke_2012}. Subsequently, the primary motivation of this work is to establish whether weak measurements induce any form of asymmetry to be introduced in an interacting network.

Our main contribution to the field is furthering the understanding of how measurement theory can be used as a mechanism of control by inducing asymmetry between two parties that have access to a shared state. In particular, this has been studied using game-theoretic techniques, which map to experimentally accessible frameworks \cite{Solmeyer_2018,PhysRevLett.88.137902}. The proposed techniques could be used to design future interacting networks.

The paper is organized as follows. In Sec.~\ref{sec:generalized_measurements}, we briefly describe the concept of generalized measurements and define the measurement operators we use in this work. In Sec.~\ref{sec:bayesian_game_theory}, we introduce the notions of Bayesian games in the context of quantum mechanics and describe which type of quantum state is used. In Sec.~\ref{ssec:measurement_induced_asymmetry}, we demonstrate a particular example of a Prisoner's dilemma, i.e., we provide a given table of payoffs and allowed measurements, that lead to asymmetry in the expected payoffs for each prisoner when the prisoners implement a particular type of generalized measurement on their share of the quantum system. We also show that one prisoner can tune their measurements to ensure one of the prisoner's expected payoffs is independent of the other prisoner's detector settings.

\section{Generalized Measurements}\label{sec:generalized_measurements}
In this section, we provide a brief summary of the concept of
weak measurements, and define the measurement operators we use throughout this paper, which, under certain limits, correspond to either weak or projective measurements. 
\subsection{Weak Measurements}
A possible way of implementing a weak measurement on a system $S$ prepared in the state $\ket{\psi} \in \mathcal{H}_S$ goes  as follows (an alternate way is shown in Ref.~\cite{PhysRevLett.95.110409}):
\begin{enumerate}[(i)]
    \item  An ancillary system $D$ initially prepared in the state $\ket{\phi} \in \mathcal{H}_D$ is coupled to $S$ via an interaction Hamiltonian
\begin{equation}\label{eq:ed0}
 H_{DS} =  \sum_{n}\sqrt{\frac{\gamma_n}{\delta t}}A_n\otimes B_n,   
\end{equation}
 where  $\delta t$ is the interaction time, and $\{\gamma_n\}$ are measurement strengths with units of inverse time such that the weak-measurement condition $\sqrt{\gamma_n \delta t} = o(1) $ holds for all $n$. Furthermore, one can also assume that the conditions $J_D/(\gamma_n \delta t) = J_S/(\gamma_n\delta t) = o(1)$ are satisfied for all $n$ so the self-evolution of both the system and detector can be ignored. Here, $J_D$ and $J_S$ are the energy scales of the detector and system Hamiltonians, respectively~\cite{attal_repeated_2006}.
 \item The total system $S+D$, which initially prepared in the state $\ket{\phi}\otimes \ket{\psi}$, evolves during a time $\delta t$ and the ancilla is \emph{projectively} measured on a given basis $\{\ket{\alpha}\}$. Therefore, the updated state of $S$ is given by 
 \begin{equation}\label{eq:ed1}
 \ket{\psi_\alpha} = \frac{M_\alpha\ket{\psi}}{\sqrt{\bra{\psi}M_\alpha^\dagger M_\alpha \ket{\psi}}},    
 \end{equation}
 where the set of Kraus operators $\{M_\alpha\}= \{\bra{\alpha}\exp(-i\delta tH_{{DS}})\ket{\phi}\}$ describes every possible indirect measurements on $S$ given the interaction \eqref{eq:ed0}, the state $\ket{\phi}$ and the measurement basis $\{\ket{\alpha}\}$. By construction, the Kraus operators satisfy $\sum_\alpha M_\alpha^\dagger M_\alpha = I_S$. 
 
 \item A post-selection of the states is performed. This step is required depending on the chosen measurement basis, as the back action of the measured ancilla can induce an effective discontinuous evolution on $\ket{\psi}$; that is, under the weak-measurement condition, upon expanding $M_\alpha $ as a power series in $\delta t$, depending on the post-selected state $\ket{\alpha}\otimes \ket{\psi_\alpha}$, the latter does not evolve continuously from the initial state $\ket{\phi}\otimes\ket{\psi}$. Therefore, the term \emph{weak} comes from the (possibly) post-selected trajectories of the effectively evolved state that evolved continuously and \emph{non-unitarily} from the prior state $\ket{\psi}$. 
\end{enumerate}

\subsection{Meausurement operators}
In what follows, we define the generalized measurements we use throughout this work and briefly outline the conditions they satisfy.

Let $\sigma, \sigma' \in \{1,- 1\},$ $\bm \alpha \in \{\bm{a},\bm{a}'\}$ and $\bm \beta \in \{\bm{b},\bm{b}'\},$ where  $\bm{a}, \bm {a}', \bm{b}, \bm{b}'$ are fixed Bloch vectors parameterized in spherical coordinates. We define the generalized measurements acting on 4-by-4 density matrices via the following measurement operators \cite{nielsen2002quantum,breuer}
\begin{equation}\label{eq:e1}
    K(\bm \alpha,\bm \beta;\sigma,\sigma';y,z) = \sqrt{P(\bm \alpha, \bm\beta)}M_{\sigma\vert \bm \alpha}(y)\otimes M_{\sigma'\vert \bm \beta}(z),
\end{equation}
where $P(\bm \alpha, \bm \beta)$ is the probability obtaining $(\bm \alpha,\bm \beta)$, and 
\begin{equation}\label{eq:3}
M_{\pm 1 \vert \bm \alpha} (y) = a(\pm y)\Pi_{+1\vert \bm \alpha}+a(\mp y)\Pi_{-1\vert \bm \alpha}
\end{equation}
are also measurement operators. Here,
\begin{equation}\label{eq:5}
\begin{split}
    \Pi_{\sigma|{\bm{\alpha}}} &= \frac{1}{2} ( \mathbb{1} +\sigma {\bm{\alpha}}\cdot{\boldsymbol{\sigma}} ) \\ &= \frac{1}{2} \begin{pmatrix} 1 + \sigma \cos \theta_{\alpha} & \sigma e^{-i\phi_{\alpha}} \sin \theta_{\alpha} \\ \sigma e^{i\phi_{\alpha}}\sin\theta_{\alpha} & 1 -\sigma \cos \theta_{\alpha} \end{pmatrix},
\end{split}
\end{equation}
is the projector associated with the Bloch vector $\bm \alpha = (\sin\theta_\alpha \cos \phi_\alpha, \sin\theta_\alpha \sin \phi_\alpha,\cos\theta_\alpha)$, and $\boldsymbol{\sigma}$ is the vector of Pauli matrices. Furthermore, 
\begin{equation}\label{eq:4}
    a(\pm y) = \sqrt{\frac{1\pm \tanh y}{2}},
\end{equation}
where $y$ characterizes the strength of the measurement. We only consider $y \in [0,\infty)$ for simplicity. For example, if $0<y \ll 1$, the measurement is denoted as \emph{weak}. Whereas if $y = 0$, then $M_{\sigma\vert \bm \alpha}(0) = I_2/\sqrt{2}$, where $I_2$ denotes the 2-by-2 identity, and it means that no measurement is performed on $\rho$.  In contrast, if $y\rightarrow \infty$, i.e., \eqref{eq:3} becomes a single projector, i.e., we obtain a strong (projective) measurement. All the above definitions also hold for $\sigma'$ and $\bm \beta$. 

For completeness, it is straightforward to verify that the measurement operators in \eqref{eq:e1} are well-defined for any $y$ and $z$: First, they resolve the identity in $\mathbb{C}^4$
\begin{equation}\label{eq:e3}
    \sum_{\bm \alpha,\bm \beta, \sigma,\sigma'}K^\dagger(\bm \alpha,\bm \beta;\sigma,\sigma';y,z)K(\bm \alpha,\bm \beta;\sigma,\sigma';y,z) = I_4,
\end{equation}
which implies 
\begin{equation}\label{eq:e4}
\sum_{\bm \alpha,\bm \beta,\sigma,\sigma'}\Tr[K(\bm \alpha,\bm \beta;\sigma,\sigma';y,z)\rho K^ \dagger(\bm \alpha,\bm \beta;\sigma,\sigma';y,z)] = 1,
\end{equation}
for an arbitrary density matrix $\rho$. Also, since we defined \eqref{eq:4} as a positive linear combination of projectors, it implies that the measurement operators are completely positive \cite{breuer}, i.e., $I_4\otimes K^\dagger(\bm \alpha,\bm \beta;\sigma,\sigma';y,z) K(\bm \alpha,\bm \beta;\sigma,\sigma';y,z) > 0$.

\section{Bayesian Game Theory}\label{sec:bayesian_game_theory}
The game-theoretic focus of this work is on Bayesian (or incomplete information) games. These were formulated classically by Harsanyi \cite{harsanyi} and have been extended in recent years to include quantum mechanics \cite{bellbayes,IqbalBayes,PhysRevA.96.042340,BayesLowe}. The essence is that players may be playing various games but are uncertain (i.e., they can have differing priors) about which game they are playing. Therefore, the probability distributions determining how the players are correlated with each other are conditioned upon which game they are playing and their subsequent choices. The conditional probability and priors determine the expected payoff for the players, where the actual payoffs can be arbitrary real numbers but are predetermined and known to both players. Importantly, there is a preference relation for these payoffs for the players associated with the game that they are playing. Throughout this paper, it is assumed that the payoff with the largest number is preferred by the players; therefore, as an example, assuming payoffs given by three and two, three is preferred over two. Subsequently, each player wishes to maximize their respective expected payoff.

In this work, we consider an incomplete information game of two players, $A$ and $B$, who share a quantum state $\rho$ describing a two-qubit system, where each  player can perform local measurements represented by \eqref{eq:3} on their subsystem. The parameters associated with players $A$ and $B$ are $\{\sigma,\bm \alpha,y\}$ and $\{\sigma',\bm \beta,z \}$, respectively, where $\bm a, \bm{a}', \bm b, \bm b'$ are fixed measurement directions, and the strengths of the measurements $y$ and $z$ are fixed throughout the game.
The incompleteness of this game is reflected in the fact that the players do not know beforehand in which direction they will be told to measure, e.g., in one trial, player $A$ can be told to measure $\bm \alpha = \bm a$ whereas player $B$ measures $\bm \beta = \bm b'$. It is an independent referee who tells each player what direction to measure. After the measurements have been performed, the results of the measurements are communicated back to the referee, and respective payoffs are assigned. An expected (or average) payoff can be calculated for each player once this procedure has been repeated many times.

Given the above settings of the game, we define the conditional probability of obtaining $(\sigma,\sigma')$ given $(\bm \alpha, \bm \beta )$ as
\begin{multline}\label{eq:1}
    P_{yz}(\sigma,\sigma' | \bm \alpha, \bm  \beta) \\ = \Tr [M_{\sigma|\bm \alpha} (y) \otimes M_{\sigma'|\bm \beta} (z)\,  \rho \,  M_{\sigma|\bm \alpha}^\dagger (y) \otimes M_{\sigma'|\bm \beta}^\dagger (z)],
\end{multline}

where the density matrix the two players share  is the specific discorded state
\begin{equation}\label{eq:12}
    \rho = \frac{1}{2}\left( \ket{00} \bra{00} + \ket{xx} \bra{xx} \right),
\end{equation}
where $\ket{x}= \cos (x/2) \ket{0} + \sin( x/2) \ket{1}$ and $\ket{0},\ket{1}$ are in the computational basis such that $\sigma^z\ket{0} = \ket{0}$. Other states are treated in Appendices~\ref{sec:app_A} and \ref{sec:app_B}.
Given the above ingredients, the expected payoff \cite{harsanyi} in our incomplete information game is given by 
\begin{multline}\label{eq:6}
    u_A(y,z;x;\theta_a, \theta_b, \theta_a', \theta_b') \\ = \sum_{\sigma,\sigma',\bm \alpha,\bm \beta}  u_{A,\sigma,\sigma'}^{\bm \alpha,\bm \beta} P(\bm \alpha, \bm \beta) P_{yz}(\sigma,\sigma' | \bm \alpha, \bm \beta),
\end{multline}
where $u_{A,\sigma,\sigma'}^{\bm \alpha,\bm \beta}$ denotes a tensor of payoffs for player $A$, $P(\bm \alpha,\bm \beta)$ denotes the prior probability, and the $\theta$'s denote the angles parameterizing the Bloch vectors associated to the measurement directions. Similar expressions hold for player $B$. 

It is important to note that in our game setup, the correlation present in the state \eqref{eq:12} can be reproduced through shared randomness by a suitable choice of a different classical correlation, as \eqref{eq:12} is a mixed, separable state. However, in the context of measurement theory, the state is quantum, so it is ultimately affected by differing measurements, which motivates the need for the following analysis.

\begin{table}
    \centering
     \begin{adjustbox}{width=0.3\textwidth}
    \begin{tabular}{c|c c}
       & Honest & Corrupt\\ \hline
    Honest  & \begin{tabular}{c|c c}
       &  {Q} &{T}\\ \hline
       {Q}  & 2,2 &0,3  \\
      {T} & 3,0 & 1,1
     \end{tabular}  & \begin{tabular}{c|c c}
       & {Q} & {T} \\ \hline
       {Q} & 2,1 & 0,0 \\
       {T} & 3,2 & 1,3
     \end{tabular} \\ 
    Corrupt  & \begin{tabular}{c|c c}
       &  {Q} & {T} \\ \hline
        {Q} &  1,2 & 2,3 \\
        {T} & 0,0 & 3,1
      \end{tabular} & \begin{tabular}{c|c c}
       &  {Q} & {T} \\ \hline
       {Q} & 1,1  & 2,0 \\
        {T} & 0,2 & 3,3
      \end{tabular}
    \end{tabular}
    \end{adjustbox}
    \caption{Bayesian version of the Prisoner's dilemma. Q represents quiet, and T represents tell. The police officer's motives and the prisoner's choices can be mapped onto ${\bm{a}},{\bm{a}'},{\bm{b}},{\bm{b}'}, +1$ and $-1$ directly in the measurement operators \eqref{eq:3}, e.g., for prisoner $A$,
 Honest $\mapsto {\bm{a}}$, Corrupt $\mapsto {\bm{a'}}$, Q $\mapsto +1$, and T $\mapsto -1$. For prisoner $B$, Honest $\mapsto {\bm{b}}$, Corrupt $\mapsto {\bm{b}'}$, Q $\mapsto +1$, and T $\mapsto -1$. Here, prisoner $A$ chooses the rows and left entries, and prisoner $B$ chooses the columns and right entries. For example, consider that prisoner $A$ believes that the police officer is honest and measures $T$, and prisoner $B$ believes that the police officer is corrupt and measures $Q$, then prisoner $A$ gets a payoff of 3, and prisoner $B$ acquires a payoff of 2.}\label{table:1}
\end{table}

\subsection{Measurement-induced asymmetry} \label{ssec:measurement_induced_asymmetry}
To clarify what asymmetry means in this context, it is instructive to understand that, by using projective measurements, the expected payoff in Bayesian game theory usually depends on both players' measurements. Therefore, if a local measurement can be chosen such that player $A$'s expected payoff becomes independent of player $B$'s measurement, it is unanticipated that player $B$'s expected payoff still depends on player $B$'s measurements. This difference in measurement dependence on the expected payoffs is defined as \emph{asymmetry moving forward} throughout this paper. This applies if players $A$ and $B$ are interchanged.

Collapsing the expected payoff so that it is independent of the other player's measurements could have practical benefits for targeting a given value. This can be illustrated by using the Prisoner's dilemma. The standard version of the game considers two prisoners who have been arrested for two offenses; one is a major offense (e.g., bank robbery), and the other is a minor offense (e.g., petty theft), but there is only enough evidence to charge them for the minor offense. The police would also like to charge them for the major offense, but there is insufficient evidence of this. Therefore, the police give the prisoners a choice where each prisoner can give evidence incriminating the other prisoner in exchange for a better prison sentence. However, if they both incriminate each other, then each prisoner will serve a longer prison sentence. Crucially, neither prisoner knows what the other will do.

In what follows, we consider a Bayesian version of the Prisoner's dilemma with the added caveat that the prisoners do not know whether their respective police officer interviewing them is either honest (H) or corrupt (C).

Since the corrupt police officer does not want to get caught, the officer rewards the prisoners for confessing and taking responsibility for the crime, i.e., tell (T), which means that C will not get caught. The reward can be considered a financial payment, but the payoffs are generally arbitrary.  However, if the prisoners remain quiet (Q), this increases the chance of C being caught, so the prisoners are not rewarded. A version of this type of game can be seen in Table~\ref{table:1}.

The expected payoffs are computed using \eqref{eq:6} with the payoffs shown in Table~\ref{table:1} for the discorded state \eqref{eq:12} and read
\begin{widetext}
\begin{multline}
\label{eq:20}
u_A(y,z;x;\theta_a,\theta_b,\theta_{a'},\theta_{b'}) \equiv u_{A}= \frac{3}{2}\\ - \frac{1}{8}\tanh(y) \Bigg\lbrace \cos \theta_a + \cos (\theta_a - x) - \cos (\frac{\theta_b - \theta_{b'}}{2}) \Bigg[ \cos (\theta_{a'} - \frac{\theta_b + \theta_{b'} }{2} )+ \cos ( \theta_{a'} - x +  \frac{\theta_b + \theta_{b'} }{2}) \cos x \Bigg] \tanh(z) \Bigg\rbrace,
\end{multline}
and
\begin{multline}
\label{eq:21}
    u_B(y,z;x;\theta_a,\theta_b,\theta_{a'},\theta_{b'}) \equiv u_{B} = \frac{3}{2}\\ - \frac{1}{8}\tanh(z) \Bigg\lbrace \cos \theta_b + \cos (\theta_b - x) - \cos (\frac{\theta_a - \theta_{a'}}{2}) \Bigg[ \cos (\theta_{b'} - \frac{(\theta_a + \theta_{a'} )}{2} ) + \cos ( \theta_{b'} - x +  \frac{\theta_a + \theta_{a'}}{2}) \cos x \Bigg] \tanh(y) \Bigg\rbrace,
\end{multline}
\end{widetext}
where $u_{A,B}$ denotes the payoffs computed for the discorded state in \eqref{eq:12} for $A$ and $B$, respectively, and $x$ is a parameter controlling the correlation. Also, it was assumed that neither prisoner $A$ nor prisoner $B$ knew whether the officer was honest or corrupt with probability 1/2. Therefore, the prior was assumed to be 1/4 as the prisoners' prior belief is statistically independent. Additionally, all $\phi$ terms were chosen to be zero as a simplification.

It can be seen that there is an asymmetry in the expected payoffs that arises due to the measurement when considering only one of the player's payoffs. For example, considering prisoner $A$'s payoff. This prisoner could set $y = 0$ (no measurement), which would collapse all the terms except $3/2$ in \eqref{eq:20}. Prisoner $B$ does not have this level of control over prisoner $A$'s payoff, as if $z =  0$, there remain other terms that Prisoner $A$ controls. Interestingly, this asymmetry in the expected payoffs is reversed when considering the other player's payoff in \eqref{eq:21}. Namely, prisoner $B$ does not have the same control over prisoner $A$'s payoff as prisoner $A$ if the former decides not to measure. 

It is clear that in the projective limit for both players, the payoffs are symmetric in the sense that $u_A$ becomes $u_B$ when changing player $A$'s measurements to player $B$'s measurements, such that $\theta_a \rightarrow \theta_b$ and so on. This is not necessarily the case when weak measurements are considered.
Despite the asymmetry in the expected payoffs due to differing types of measurement, there is still the possibility of a given prisoner optimizing the respective payoff for a given choice of measurement. 
For example, prisoner $A$ can choose to make no measurement ($y=0$) and fix the expected payoff at $3/2$. Despite this, prisoner $B$ would still be left to control the measurements for the payoff $u_B$. Thus, in theory, prisoner B would be able to acquire a larger expected payoff than prisoner $A$ by choosing the detector settings ($\theta_b, \theta_{b'}$) in an optimal way and vice versa for prisoner $B$. Subsequently, each player (if they so choose) has complete control over their respective expected payoff and can make their expected payoff completely independent of the other player's choices (measurements) while allowing the other player to still have control over the corresponding particular payoff. This could have an application where one player is more risk-averse compared to the other, so this player can choose whether to allow the other player to influence their respective payoff.

\subsection{Measurement-induced control}\label{ssec:measurement_induced_control}

This section is devoted to determining whether there is an inherent advantage when one prisoner is restricted to a given type of measurement. Initially, suppose one prisoner is constrained to projective measurements and the other one is restricted to weak measurements. This type of network could be predetermined by the referee who controls the network. Given these restrictions, the prisoners would have to determine the optimal strategies in order to maximize their respective payoffs. At this stage, it is worth emphasizing that the players know the payoffs and the correlation between them so they can compute their respective expected payoffs and make optimal strategic decisions based on the known information. 

By considering the expressions above in \eqref{eq:20} and \eqref{eq:21}, and limiting prisoner $B$'s measurement to be projective ($z\rightarrow \infty$ or  $\tanh z \rightarrow 1$), one can see that this restriction does not offer any advantage, in the sense that both players still have full control over their respective payoffs through measurement. Subsequently, standard game-theoretic techniques would still be required to find the optimal solution (i.e., Nash equilibrium \cite{540b73bd-a3f1-333e-a206-c24d0fbbb8bc}). Similar expressions would be calculated if prisoner $A$ was restricted to projective measurements instead of $B$.

\begin{figure}
    \includegraphics[scale=0.55]{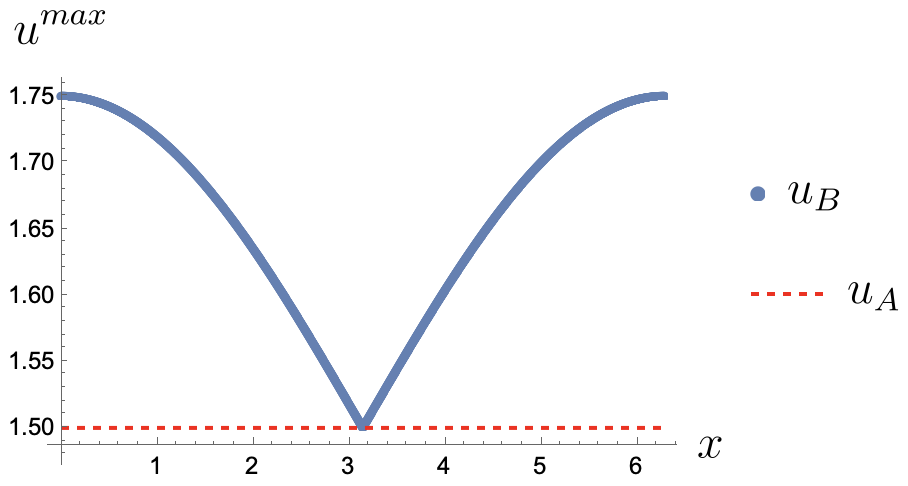}
    \makeatletter\long\def\@ifdim#1#2#3{#2}\makeatother
    \caption{This figure shows the maximum expected payoff for both players as a function of $x$. Prisoner $B$ can choose the measurement settings ($\theta_{b}$) in order to maximize the expected payoff $u_B$ for each value of $x$. Therefore, this measurement-induced asymmetry gives an inherent advantage to prisoner $B$. It is seen that the extrema occur at the classical correlation of the state, namely $x=0,n\pi$ where $n \in \mathbb{Z}$. }
    \label{maxPayoff}\label{fig:1}
\end{figure}

On top of the strong measurement performed by prisoner $B$, let us assume that prisoner $A$ is restricted to a weak measurement, which, up to a first-order series expansion around $y = 0$, yields
\begin{widetext}
\begin{multline}
\label{eq:weak}
u_A(y,z;x;\theta_a,\theta_b,\theta_{a'},\theta_{b'}) \equiv u_{A}= \frac{3}{2}\\ - \frac{1}{8} y \Bigg\lbrace \cos \theta_a + \cos (\theta_a - x) - \cos (\frac{\theta_b - \theta_{b'}}{2}) \Bigg[ \cos (\theta_{a'} - \frac{\theta_b + \theta_{b'} }{2} )+ \cos ( \theta_{a'} - x +  \frac{\theta_b + \theta_{b'} }{2}) \cos x \Bigg] \Bigg\rbrace,
\end{multline}
and
\begin{multline}
\label{eq:weak1}
    u_B(y,z;x;\theta_a,\theta_b,\theta_{a'},\theta_{b'}) \equiv u_{B} = \frac{3}{2}\\ - \frac{1}{8} \Bigg\lbrace \cos \theta_b + \cos (\theta_b - x) - \cos (\frac{\theta_a - \theta_{a'}}{2}) \Bigg[ \cos (\theta_{b'} - \frac{(\theta_a + \theta_{a'} )}{2} ) + \cos ( \theta_{b'} - x +  \frac{\theta_a + \theta_{a'}}{2}) \cos x \Bigg] y \Bigg\rbrace.
\end{multline}
\end{widetext}
It is clear that in the weak measurement limit, prisoner $A$ has differing levels of control over each payoff function. To explicitly highlight this, if we take the extreme case of no measurement ($y=0$), then \eqref{eq:weak} and \eqref{eq:weak1} become
\begin{equation}
\label{eq:24}
    u_{A} = \frac{3}{2}
\end{equation}
and 
\begin{equation}
\label{eq:25}
\begin{split}
    u_{B} = \frac{3}{2} - \frac{1}{8} \Big[ \cos \theta_b + \cos (\theta_b - x) \Big],
\end{split}
\end{equation}
respectively.

Now, it is seen that prisoner $B$ can choose particular measurement directions to optimize the payoff $u_B$. It is also interesting to note that this depends on the correlation controlled by $x$. One solution for the maximum of $u_{B}$ would be when $\theta_b = \pi$ and $x=0$. This would yield a result of $u_{B} = 7/4$. Therefore, it is in prisoner $B$'s interest for prisoner $A$ to be restricted to a no-measurement, as this allows prisoner $B$ to maximize $u_B$, ensuring it is larger than $3/2$. One can envisage scenarios where the effect of making another party's operations redundant would have significant practical benefits in the design of future interacting networks, where one party having control over the network is a necessity of the design. Subsequently, we expect measurement theory to be used as a mechanism for the future design of secure quantum networks.  Figure~\ref{fig:1} shows the maximum expected payoff for both players as a function of $x$, where the maximization is taken over $\theta_b$ for each value of $x$ when $z \rightarrow \infty$ and $y  = 0$.

\section{Discussion}
There are a couple of key comments that are important for this work. While Bayesian games are considered to see how two parties can interact across a network, determining the Nash equilibrium was not been considered. This is because the focus of the work is revealing the measurement-induced asymmetry. It is an interesting future question to understand how this asymmetry impacts Nash equilibria and whether the measurement process allows increasingly optimal solutions. 

It is clear (compare Appendix~\ref{sec:app_A} with Appendix~\ref{sec:app_B}) that the measurement-induced asymmetry is also strongly related to the choice of payoffs, suggesting the game-theoretic nature of this work. In fact, there is a general relation between the choice of payoffs and witnessing measurement-induced asymmetry as for an arbitrary two qubit state, the functional form of the expected payoff is the same, thus asymmetry is always present, but may only be revealed for certain choices of payoffs. 

Moreover, while a discorded state was used to demonstrate the measurement-induced asymmetry in \eqref{eq:weak} and \eqref{eq:weak1}, the state itself was symmetric because the quantum discord is the same irrespective of which subsystem the measurement is performed on. Therefore, this further justifies the claim that the asymmetry is entirely due to the measurement process. In the future, studying how an asymmetric (one-way discorded) state affects the measurement will also be of practical relevance.

Since only bipartite networks have been considered, a natural extension is to include further parties in the interaction and determine the effect of weak measurements for this extended network. There is also the question of the specific network topology for an increasing number of parties. Determining the relation between network topology and weak measurement is another interesting avenue to explore.

Furthermore, determining the general relation between generalized measurements and correlations that reveal asymmetry is the key future question this research raises.

The experimental realization of this work in terms of generalized measurements is a feasible task since weak measurements have been performed in, e.g., superconducting qubits \cite{qubit_quantum_traj_exp,quantum_trajectory_tracking_nn}. Moreover, the bipartite network we have formulated in this work can be readily implemented in an optical experimental setup. For example, in Ref.~\cite{Choi:20}, the measurement operators \eqref{eq:3} were used.

\section{Summary}
In this paper, we have established that generalized measurements naturally introduce asymmetry in a bipartite network given a discorded state when one party is restricted to projective measurements. Importantly, we have shown that this asymmetry allows a given party to control the other party's expected payoff through the choice of measurements, as expected payoffs can be made independent of the detector settings. This could be used as a feasible technique in future quantum network design for designing systems with inherent control to given nodes on the network.

\section{Acknowledgements}
AL is grateful for hospitality at Weizmann Institute and acknowledges funding provided by the Department for Science, Innovation and Technology (DSIT) via the Universities “UK-Israel innovation researcher mobility scheme” grant number 1004. AL also acknowledges stimulating discussions with Azhar Iqbal. EMG acknowledges the Grants No.~EG 96/13-1 and No.~GO 1405/6-1. In addition, we acknowledge
funding by the Israel Science Foundation

\appendix

\section{CHSH Game}\label{sec:app_A}

In order to verify and confirm the analysis conducted in this paper, it is prudent to reproduce known results within the expected limits. Moreover, this will help clarify whether measurement-induced asymmetry is a purely quantum process or whether it generalizes to non-quantum correlations.

Previous work has been done in this setting using a Bell state \cite{PhysicsPhysiqueFizika.1.195} given by
\begin{equation}\label{eq:9}
    \rho_Q = \ket{\psi} \bra{\psi}, \hspace{0.5cm} \ket{\psi} = \frac{1}{2} \left( \ket{00} + \ket{11} \right).
\end{equation}
Furthermore, an interesting extension of the Bell state is to consider the Werner state \cite{PhysRevA.40.4277}. This is because the Werner state's correlation evolves from classically uncorrelated to discorded, entangled, and non-separable as a function of a parameter $\eta$. It is defined by
\begin{equation}\label{eq:11}
    \rho_W = \frac{1-\eta}{4}I_4 + \eta \ket{\phi}\bra{\phi},
\end{equation}
where $\ket{\phi}= \dfrac{1}{\sqrt{2}}\big[ \ket{01} - \ket{10} \big]$. 

\begin{table}
    \centering
     \begin{adjustbox}{width=0.3\textwidth}
    \begin{tabular}{c|c c}
     ($\bm \alpha,\bm \beta$)   & ${\bm{b}}$ &${\bm{b'}}$ \\ \hline
     ${\bm{a}}$  & \begin{tabular}{c|c c}
       &  {$\uparrow$} &{$\downarrow$}\\ \hline
       {$\uparrow$}  & 1,-1 &0,0  \\
      {$\downarrow$} & 0,0 & 1,-1
     \end{tabular}  & \begin{tabular}{c|c c}
       & {$\uparrow$} & {$\downarrow$} \\ \hline
       {$\uparrow$} & 1,-1 &0 ,0 \\
       {$\downarrow$} & 0,0 & 1,-1
     \end{tabular} \\ 
     ${\bm{a'}}$  & \begin{tabular}{c|c c}
       &  {$\uparrow$} & {$\downarrow$} \\ \hline
        {$\uparrow$} &  1,-1 &0,0 \\
        {$\downarrow$} & 0,0 & 1,-1
      \end{tabular} & \begin{tabular}{c|c c}
       &  {$\uparrow$} & {$\downarrow$} \\ \hline
       {$\uparrow$} & 0,0  & 1,-1 \\
        {$\downarrow$} & 1,-1 & 0,0
      \end{tabular}
    \end{tabular}
    \end{adjustbox}
    \caption{This table shows the payoff matrix for a zero-sum CHSH game. The first number in each element is the payoff to player A (Alice), and the second number is the payoff to player B (Bob). In the traditional CHSH game, Alice's and Bob's payoffs are equal and are the first numbers in each element. Based on what bits they receive, ${\bm{a}},{\bm{b}},{\bm{a}}'$, and ${\bm{b}}'$ denote how Alice and Bob's detectors are set up. From this, they then perform measurements on their shared states, and based on their results, they are assigned a payoff. Also,  $\bm \alpha\in \{{\bm{a}},{\bm{a'}}\}$, $\beta\in \{{\bm{b}},{\bm{b'}}\}$ and therefore denotes what bits the players receive either 0 or 1. This is analogous to Table~\ref{table:1}.}\label{table:2}
\end{table}

It is important to consider a known example to verify that the framework with generalized measurements produces the expected results in the projective measurement limit. Therefore, by using the CHSH game \cite{1313847}, which is a Bayesian game-theoretic version of the CHSH inequality \cite{PhysRevLett.23.880}, it can be established that the correct results are obtained, specifically for the Bell state. The standard CHSH game is found by considering player $A$'s expected payoffs given in Table~\ref{table:2}.

Using the Bell state defined in \eqref{eq:9} yields an expected payoff of 
\begin{equation}
\begin{split}
    u_{A}^{Q} &= \frac{1}{8} \Big[ 4 + \Big( \cos(\theta_{\bm{a}} - \theta_{\bm{b}})+\cos(\theta_{\bm{a'}} - \theta_{\bm{b}}) \\&+ \cos(\theta_{\bm{a}} - \theta_{\bm{b'}}) -\cos(\theta_{\bm{a'}} - \theta_{\bm{b'}}) \Big) \\& \times \tanh(y) \tanh(z) \Big]. 
\end{split}
\end{equation}
Since it is a zero-sum game, player $B$'s expected payoff is the negative of player $B$'s $(u_{Q}^{B} = - u_{A}^{Q})$.
Taking $y\rightarrow \infty$ reproduces the expected result for the CHSH game \cite{1313847}. Furthermore, maximizing the resultant function yields $(1/8)(4 + 2 \sqrt{2}) \approx 0.85$, which is consistent with known results. It is interesting to note the maximum occurs for a projective measurement for both players. However, either player can choose to collapse the cosine terms such that $y= z =  0 $ gives $u_A = 1/2$. 

For the discorded state in \eqref{eq:12}, the expected payoff is
\begin{widetext}
\begin{multline}
    u_{A}^{D} = \frac{1}{32} \Big\{ 16 + \Big[ 2\Big( \cos(\theta_{\bm{a}}-\theta_{\bm{b}}) + \cos(\theta_{\bm{a'}}-\theta_{\bm{b}})  + 
    \cos(\theta_{\bm{a}}-\theta_{\bm{b'}}) - \cos(\theta_{\bm{a'}}-\theta_{\bm{b'}}) \Big) +  \cos(\theta_{\bm{a}}+\theta_{\bm{b}}) +  \cos(\theta_{\bm{a'}}+\theta_{\bm{b}}) \\+  \cos(\theta_{\bm{a}}+\theta_{\bm{b'}}) - \cos(\theta_{\bm{a'}}+\theta_{\bm{b'}})+ 
     \cos(\theta_{\bm{a}}+\theta_{\bm{b}}-2x) +  \cos(\theta_{\bm{a'}}+\theta_{\bm{b}} -2x)  +  \cos(\theta_{\bm{a}}+\theta_{\bm{b'}} -2x) \\- \cos(\theta_{\bm{a'}}+\theta_{\bm{b'}} -2x) \Big] \tanh(y) \tanh(z) \Big\}.
\end{multline}
\end{widetext}
Taking the projective measurement limit and maximizing over all variables gives a maximum payoff of $0.75$, which is the classical limit. The discorded state is local, whereas the Bell state is non-local. Subsequently, it is clear that the expected results are reproduced for this framework.
The expected payoff for the discorded state is similar in behavior to the corresponding payoff for the Bell state in the sense that both players have the same control over the expected payoff due to their respective weak measurements. Additionally, the expected payoff for the Werner state is similar to the Bell state, except for the coupling parameter $\eta$ and a change of sign, which results in 
\begin{equation}
\begin{split}
    u_{A}^{W} &= \frac{1}{8} \Big[ 4 - \eta \Big( \cos(\theta_{\bm{a}} - \theta_{\bm{b}}) +\cos(\theta_{\bm{a'}}  - \theta_{\bm{b}}) \\ &\,\,+\cos(\theta_{\bm{a}} - \theta_{\bm{b'}})-\cos(\theta_{\bm{a'}} - \theta_{\bm{b'}}) \Big) \\ &\,\,\times \tanh(y) \tanh(z) \Big]. 
\end{split}
\end{equation}
Similar results can be reproduced for this state. The uncorrelated classical state has an expected payoff of $1/2$ for player A and $-1/2$ for player B, and the quantum maximum can be achieved for a suitable choice of $\eta$.

\section{Modified CHSH Game}\label{sec:app_B}
Let us consider a modified combative CHSH game given in Table~\ref{table:3}. Notice how it is still a zero-sum game. However, the payoff matrices for each type of game are no longer symmetric along the diagonal elements. Therefore, there is an added complication which the players need to take into account. However, it is worth emphasizing that the only modification is the payoffs.
\begin{table}
    \centering
     \begin{adjustbox}{width=0.3\textwidth}
    \begin{tabular}{c|c c}
     ($\bm \alpha,\bm \beta$)   & ${\bm{b}}$ &${\bm{b'}}$ \\ \hline
     ${\bm{a}}$  & \begin{tabular}{c|c c}
       &  {$\uparrow$} &{$\downarrow$}\\ \hline
       {$\uparrow$}  & 1,-1 &1,-1  \\
      {$\downarrow$} & -1,1 & -1,1
     \end{tabular}  & \begin{tabular}{c|c c}
       & {$\uparrow$} & {$\downarrow$} \\ \hline
       {$\uparrow$} & 1,-1 & 1,-1 \\
       {$\downarrow$} & -1,1 & -1,1
     \end{tabular} \\ 
     ${\bm{a'}}$  & \begin{tabular}{c|c c}
       &  {$\uparrow$} & {$\downarrow$} \\ \hline
        {$\uparrow$} &  1,-1 & 1,-1 \\
        {$\downarrow$} & -1,1 & -1,1
      \end{tabular} & \begin{tabular}{c|c c}
       &  {$\uparrow$} & {$\downarrow$} \\ \hline
       {$\uparrow$} & -1,1  & 1,-1 \\
        {$\downarrow$} & 1,-1 & -1,1
      \end{tabular}
    \end{tabular}
    \end{adjustbox}
    \caption{This table shows a different payoff matrix for a modified CHSH game. This game allows inherent asymmetry to be introduced due to the measurement process. The experimental setup is the same as in Table~\ref{table:2}. Note how the payoffs are no longer symmetric along the diagonal elements of the individual payoff matrices. }\label{table:3}
\end{table}

Using the payoffs in Table~\ref{table:3} and performing a similar analysis as for Table~\ref{table:2}, the Bell state results in a similar payoff because the measurement is equivalent for both players. This implies that each player has the same amount of control based on their measurement over the expected payoff. This can be seen by
\begin{equation}
    u_{A}^{Q} = -\frac{1}{4} \cos(\theta_{\bm{a'}} - \theta_{\bm{b'}}) \tanh(y) \tanh(z).
\end{equation}
Similarly, for the Werner state the result is
\begin{equation}
    u_{A}^{W} = \frac{\eta}{4} \cos(\theta_{\bm{a'}} - \theta_{\bm{b'}}) \tanh(y) \tanh(z).
\end{equation}

However, some asymmetry (using the definition given in Sec. \ref{ssec:measurement_induced_asymmetry}) is introduced entirely due to the weak measurement for the discorded state. This confirms what is found in the main text. Explicitly, this means one player has more control over the outcome of the expected payoff than the other, as if $y \rightarrow \infty$, then the expected payoff is zero for both players. However, if $z\rightarrow 0$, the expected payoffs still depend on the measurements. \textcolor{red}{} This can be seen in the expected payoffs given by
\begin{equation}\label{eq:18}
\begin{split}
   u_{A}^{D} &= \frac{1}{8} \tanh(y) \Big\{ 2 \big[ \cos \theta_{\bm{a}} + \cos(\theta_{\bm{a}} -x) \big] \\ &+ \cos \theta_{\bm{a'}}+
   \cos ( \theta_{\bm{a'}} - x) - \big[\cos(\theta_{\bm{a'}} -\theta_{\bm{b'}}) \\ &+ \cos(\theta_{\bm{a'}} +\theta_{\bm{b'}} - x) \cos x \big] \tanh(z) \Big\},
\end{split}
\end{equation}
with $u_{B}^{D} = - u_{A}^{D}$.

This is further evidence of the main claim of measurement-induced asymmetry in the main text, as the level of control over the expected payoffs differs for each player depending on what measurement is performed. 

Therefore, it is clear that the observed asymmetric effect is robust against changes in the payoffs, ensuring the measurement-induced asymmetry persists, particularly when utilizing weak measurements. This has been checked for a variety of well-known quantum states, and the observed phenomena has been confirmed in each of the cases studied. 

As a final check of the robustness of the asymmetry, an arbitrary two-qubit state was taken to determine if the same functional forms found throughout this paper are reproduced for the various payoff matrices considered. It was found for each payoff matrix, the functional form was the same. Therefore, the amount of control due to the measurement is consistent, irrespective of which state is used. This further highlights the observed effects are entirely due to measurement. In particular, we have demonstrated that weak measurement can be used as a mechanism for control and advantage when parties are performing local measurements on a bipartite network.

\bibliographystyle{apsrev4-2}
%


\end{document}